\documentstyle[12pt]{article}
\renewcommand{\thesection}{\Roman{section}}

\begin{document}
\setlength{\oddsidemargin}{-1cm}
\setlength{\evensidemargin}{-1cm}
\newcommand{\nc}{\newcommand}
\nc{\beq}{\begin{equation}}
\nc{\eeq}{\end{equation}}
\nc{\bea}{\begin{eqnarray}}
\nc{\eea}{\end{eqnarray}}
\nc{\bpi}{\begin{picture}}
\nc{\epi}{\end{picture}}
\nc{\ba}{\begin{array}}
\nc{\ea}{\end{array}}
\nc{\nn}{\nonumber}
\nc{\ts}{\textstyle}
\nc{\ds}{\displaystyle}
\nc{\hh}{\;\;}
\nc{\hf}{\;\;\;\;}
\nc{\hs}{\;\;\;\;\;\;}
\nc{\he}{\;\;\;\;\;\;\;\;}
\nc{\vs}{\vspace{6pt}}
\nc{\fm}{\frac{m^2}{\bar{\mu}^2}}
\nc{\lnma}{\ln\fm}
\nc{\lnmb}{\ln^2\fm}
\nc{\lnmc}{\ln^3\fm}
\nc{\lnmd}{\ln^4\fm}
\nc{\mb}{\bar{m}_B^2}
\nc{\msbar}{$\overline{\rm MS}$}
\nc{\twocrdg}{{\ba{c}\mbox{\small 2 crossed}\\
\mbox{\small diagrams}\ea}}
\nc{\fivecrdg}{{\ba{c}\mbox{\small 5 crossed}\\
\mbox{\small diagrams}\ea}}
\nc{\od}{{\cal O}}
\nc{\al}{\alpha}
\nc{\be}{\beta}
\nc{\ga}{\gamma}
\nc{\Ga}{\Gamma}
\nc{\de}{\delta}
\nc{\ep}{\epsilon}
\nc{\th}{\theta}
\nc{\ze}{\zeta}
\nc{\p}{\partial}

\begin{titlepage}
\begin{flushright} PURD-TH-96-03\\ hep-ph/9604311\\ April 1996
\end{flushright}
\vspace{0.4cm}
\begin{center}
{\large\bf  Four-Loop Vacuum Energy Beta Function in\\
\mbox{\boldmath O($N$)} Symmetric Scalar Theory
\vspace{0.4cm}\vspace{0.8cm}\\}
Boris Kastening\vspace{0.8cm}\\
{\it Department of Physics\\ Purdue University\\ West Lafayette,
Indiana 47907-1396}\vspace{0.8cm}\vspace{0.4cm}\\
{\bf Abstract}\\
\end{center}

\begin{center}
\begin{minipage}{16cm}
\setlength{\baselineskip}{20pt}
The beta function of the vacuum energy density is computed at the
four-loop level in massive O($N$) symmetric $\phi^4$ theory.
Dimensional regularization is used in conjunction with the \msbar\
scheme and all calculations are in momentum space in the massive
theory.
The result is $\be_v=\frac{N}{4}g+\frac{N(N+2)}{96}g^3
+\frac{N(N+2)(N+8)[12\ze(3)-25]}{1296}g^4+\od(g^5)$.
\end{minipage}
\end{center}

\end{titlepage}

\section{Introduction}

The beta function for the coupling constant, $\be_g$, the gamma
function for the mass parameter, $\ga_m$, and the anomalous dimension
$\ga_\phi$ are known to five loops \cite{Kl..} in O($N$) symmetric
$\phi^4$ theory in dimensional regularization (DR) in conjunction
with the modified minimal subtraction scheme (\msbar).
There is, however, another, less well-known, beta function $\be_v$
related to the vacuum energy density.
It was introduced in \cite{Ba..} to facilitate renormalization
group improvement of effective potentials in massive theories,
which was first performed correctly, but in a less elegant scheme,
in \cite{Ka}.
Since then it has become a standard tool for investigations of
vacuum stability in massive theories.
While in flat space $\be_v$ is more a tool of calculational
convenience, in curved spacetime it describes the running of the
cosmological constant \cite{El..1}.
$\be_v$ has never been computed to higher-loop orders in any model.
In this paper, we compute the vacuum energy beta function to four
loops in O($N$) symmetric $\phi^4$ theory, using DR and \msbar.
To our knowledge, the highest order to which $\be_v$ has been
computed in this model is one loop \cite{El..2}.

There are at least two other motivations for computing $\be_v$ to
high-loop orders.

First, there have been recent claims about a connection between
divergences in field theory and invariants of knot theory
\cite{Kr..}.
Since in any given loop order, there are considerably less vacuum
diagrams to compute than diagrams for two- and four-point functions,
this may be an easier way of tracking the connection between field
theory and knot theory.
In fact, after absorbing the one-loop mass correction into a modified
bare mass, there is only one diagram to compute in each one- to
four-loop order in the $\phi^4$ model.
At five loops there are three, at six loops six diagrams.

Second, when computing the vacuum energy beta function to four loops,
the postulate that subdivergences are cancelled by the appropriate
mass and coupling constant counterterms allows us to get as a
byproduct $\ga_m$ at three loops and $\be_g$ at two loops.
If one can make this work to higher orders, the rule will be:
Computation of $\be_v$ to $n$ loops provides $\ga_m$ at $(n-1)$
loops and $\be_g$ at $(n-2)$ loops.

It is not clear to the present author if there is any connection
to the critical theory in three dimensions via the $\ep$ expansion.

The structure of the paper is as follows.
In Section \ref{defconv} our conventions are established.
In Section \ref{relations} we provide the relations between
the $\be$ and $\ga$ functions on the one hand and the renormalization
constants $Z_x$ ($x=g,m,\phi,v$) on the other hand and also give
recursion relations for the components of the $Z_x$.
$Z_g$ and $Z_m$ are formally reconstructed from $\be_g$ and $\ga_m$
at two and three loops, respectively.
In Section \ref{oneloopresummation} the one-loop mass correction
is absorbed into a modified bare mass to significantly reduce the
number of vacuum diagrams to be computed.
In Section \ref{fourloop} we finally set out to determine the
vacuum energy density counterterms and $\be_v$ at the four-loop
level, recovering at the same time the two-loop $\be_g$ and
three-loop $\ga_m$. 
The appendices are reserved for the computation of the necessary
integrals.

\section{Definitions and Conventions}
%====================================
\label{defconv}

We work with the same conventions (except their $Z_2$, $\ga_2$
are our $Z_\phi$, $\ga_\phi$ and their $Z_{m^2}$ is our $Z_m$) as
\cite{Kl..}, but extend the usual Lagrange density by a constant
term,
\beq
\label{Lb}
{\cal L}
=
\frac{1}{2}\p_\mu\phi_{Ba}\p_\mu\phi_{Ba}
+\frac{1}{2}m_B^2\phi_B^2
+\frac{1}{24}(4\pi)^2 g_B(\phi_B^2)^2
+\frac{m_B^4 h_B}{(4\pi)^2 g_B}\,,
\eeq
where $\phi_B^2\equiv\phi_{Ba}\phi_{Ba}$, repeated indices are summed 
over ($\mu=1,\ldots,d$, $a=1,\ldots,N$) and the subscript $B$ refers
to bare quantities.
We work in $(d=4-\ep)$-dimensional Euclidean space and use DR with
\msbar.
All our loop integrations are in momentum space in the massive
theory.
The connection between bare and renormalized quantities is given by
\beq
\label{br}
g_B = \mu^\ep Z_g g\,,\he
m_B^2 = Z_m m^2\,,\he
\phi_B^2 = \mu^{-\ep} Z_\phi \phi^2\,,\he
h_B = Z_h h\,,
\eeq
where $\mu$ is the renormalization scale [connected to the \msbar\
renormalization scale by $\mu^2=\bar{\mu}^2e^{\ga_E}/(4\pi)$]
and the $Z$'s have the form
\bea
\label{zs}
\begin{array}{ll}
\ds Z_g = 1+\sum_{k=1}^\infty\frac{Z_{g,k}(g)}{\ep^k}\,,\he
&
\ds Z_{g,k}(g) = \sum_{l=k}^\infty Z_{kl}^g g^l\,,
\\&\\
\ds Z_m = 1+\sum_{k=1}^\infty\frac{Z_{m,k}(g)}{\ep^k}\,,\he
&
\ds Z_{m,k}(g) = \sum_{l=k}^\infty Z_{kl}^m g^l\,,
\\&\\
\ds Z_\phi = 1+\sum_{k=1}^\infty\frac{Z_{\phi,k}(g)}{\ep^k}\,,\he
&
\ds Z_{\phi,k}(g) = \sum_{l=k}^\infty Z_{kl}^\phi g^l\,,
\\&\\
\ds Z_v\equiv \frac{Z_m^2 Z_h}{Z_g}
  = 1+\frac{1}{h}\sum_{k=1}^\infty\frac{Z_{v,k}(g)}{\ep^k}\,,\he
&
\ds Z_{v,k}(g) = \sum_{l=k}^\infty Z_{kl}^v g^l\,.
\end{array}
\eea

There are different ways to construct the Feynman rules as far as
the treatment of counterterms is concerned.
For our purposes it is most convenient to choose
\beq
\label{feynman rules}
\ba{ll}
{\rm \underline{propagator:}}
\\
\hh
\bpi(50,12)
\put(5,0){$a$}
\put(39,0){$b$}
\put(15,3){\line(1,0){20}}
\epi
&\ds
=\frac{\de_{ab}Z_\phi^{-1}}{p^2+m_B^2}
\vs\\
{\rm \underline{vertices:}}
\\
\hh
\bpi(50,12)
\put(25,3){\circle*{4}}
\epi
&\ds
= -\frac{m_B^4 h_B}{(4\pi)^2\mu^{-\ep}g_B}
\vs\\
\hh
\rule[-10pt]{0pt}{28pt}
\bpi(50,12)
\put(5,10){$a$}
\put(39,10){$b$}
\put(5,-10){$c$}
\put(39,-10){$d$}
\put(15,-7){\line(1,1){20}}
\put(15,13){\line(1,-1){20}}
\put(25,3){\circle*{4}}
\epi
\hh
&\ds
=-[\de_{ab}\de_{cd}+\de_{ac}\de_{bd}+\de_{ad}\de_{bc}]
Z_\phi^2\,\frac{(4\pi)^2\mu^{-\ep}g_B}{3}\,.
\\
\ea
\eeq
Then no extra ``countervertices'' have to be considered.
After computing a certain diagram, the result has to be reexpanded
to the desired order in $g$.
Since we are only interested in graphs without external legs,
the wave function renormalization counterterms contained in $Z_\phi$
cancel from the outset since the total power of $Z_\phi$ for vacuum
graphs is zero.
We take the integration measure to be
\beq
\label{integration measure}
\int_p\equiv\mu^\ep\int\frac{d^d p}{(2\pi)^d}\,,
\eeq
so that all Feynman diagrams have integer dimension even for
$\ep\neq 0$.

Next we define the various $\be$ and $\ga$ functions in arbitrary
dimension:
\beq
\label{betagamma}
\ba{llll}
\ds \be_{g,\ep}(g,\ep)
&\!\!\ds =\mu^2
\left(\frac{\p g}{\p\mu^2}\right)_{\! B}\,,\he
&\ds \ga_{m,\ep}(g,\ep)
&\!\!\ds =\frac{\mu^2}{m^2}
\left(\frac{\p m^2}{\p\mu^2}\right)_{\! B}\,,
\vs\\
\ds \ga_{\phi,\ep}(g,\ep)
&\!\!\ds =-\frac{\mu^2}{\phi^2}
\left(\frac{\p \phi^2}{\p\mu^2}\right)_{\! B}\,,\he
&\ds \be_{v,\ep}(g,\ep)
&\!\!\ds =\frac{g\mu^{2+\ep}}{m^4}
\left[\frac{\p}{\p\mu^{2}}
\left(\frac{m^4 h}{\mu^\ep g}\right)\right]_{\! B}\,.
\ea
\eeq

As an aside let us mention that then the effective potential
$V_\ep$ in $4-\ep$ dimensions obeys the renormalization group
equation
\beq
\left\{
\mu^2\frac{\p}{\p\mu^2}
{+}\be_{g,\ep}\frac{\p}{\p g}
{+}\ga_{m,\ep}m^2\frac{\p}{\p m^2}
{-}\ga_{\phi,\ep}\phi^2\frac{\p}{\p \phi^2}
{+}\left[\be_{v,\ep}{-}h
\left({-}\frac{\ep}{2}{+}2\ga_{m,\ep}{-}
\frac{\be_{g,\ep}}{g}\right)\right]\frac{\p}{\p h}
\right\}V_\ep=0\,.
\eeq
Since the only term in $V_\ep$ containing $h$ is
$m^4 h/[(4\pi)^2\mu^\ep g]$, the last equation can be written as
\beq
\left[
\mu^2\frac{\p}{\p\mu^2}
+\be_{g,\ep}\frac{\p}{\p g}
+\ga_{m,\ep}m^2\frac{\p}{\p m^2}
-\ga_{\phi,\ep}\phi^2\frac{\p}{\p \phi^2}
\right]V_\ep(g,m^2,\phi^2,h=0,\mu^2)
=
-\be_{v,\ep}\frac{m^4}{(4\pi)^2\mu^\ep g}\,.
\eeq
Note that only the $\phi$-independent part of $V_\ep(h=0)$ is
affected by the inhomogeneous term.
Therefore one can get around introducing a constant term into the
Lagrange density and using $\be_v$ when renormalization group
improving the effective potential by considering
$V_\ep(\phi^2)-V_\ep(\phi_0^2)$ where $\p V_\ep/\p\phi=0$ at $\phi_0$
or by improving $\p V_\ep/\p\phi$ or $V_\ep(\phi^2)'$, since these
quantities obey the usual homogeneous renormalization group equation
(see \cite{Ka,Fo..}).
However, those methods are less elegant.

\section{Relations for the $\be_x$, $\ga_x$ and $Z_x$}
%=====================================================
\label{relations}

Using standard methods \cite{CoMa} one arrives at
\beq
\label{begares}
\ba{c}
\be_{g,\ep}(g,\ep)=-\frac{1}{2}\ep g+\be_{g}(g)\,,\he
\be_{g}(g)=\frac{1}{2}g^2 Z_{g,1}'\,,
\vs\\
\ga_{m,\ep}(g,\ep)=\ga_{m}(g)=\frac{1}{2}g Z_{m,1}'\,,
\vs\\
\ga_{\phi,\ep}(g,\ep)=-\frac{1}{2}\ep+\ga_{\phi}(g)\,,\he
\ga_{\phi}(g)=-\frac{1}{2}g Z_{\phi,1}'\,,
\vs\\
\be_{v,\ep}(g,\ep)=\be_v(g)=\frac{1}{2}g Z_{v,1}'\,,
\ea
\eeq
where the functions without index $\ep$ are the ones for the 
four-dimensional theory, i.e., $\ep\rightarrow 0$.
Therefore the $\be$ and $\ga$ functions have the simple structure
\beq
\label{betagammapowerseries}
\be_g = \sum_{k=1}^\infty\be_k g^{k+1}\,,\hf
\ga_m = \sum_{k=1}^\infty\al_k g^{k}\,,\hf
\ga_\phi = \sum_{k=1}^\infty\ga_k g^{k}\,,\hf
\be_v = \sum_{k=1}^\infty\de_k g^{k}\,,
\eeq
where the $\be_k$, $\al_k$, $\ga_k$ and $\de_k$ are just real
numbers.
In the course of deriving the relations (\ref{begares}) one can also
extract the recursion relations
\beq
\label{zrecursion}
\ba{l}
Z_{g,k+1}'=Z_{g,1}'(gZ_{g,k})'\,,
\vs\\
Z_{m,k+1}'=Z_{m,1}'Z_{m,k}+g Z_{g,1}'Z_{m,k}'\,,
\vs\\
Z_{\phi,k+1}'=Z_{\phi,1}'Z_{\phi,k}+g Z_{g,1}'Z_{\phi,k}'\,,
\vs\\
Z_{v,k+1}'=(2Z_{m,1}'-Z_{g,1}')Z_{v,k}+g Z_{g,1}'Z_{v,k}'\,,
\ea
\eeq
valid for $k\geq 1$.
We use the first two relations in both (\ref{begares}) and
(\ref{zrecursion}) together with (\ref{zs}) and
(\ref{betagammapowerseries}) to formally reconstruct the coupling
constant and mass counterterms from $\be_g$ and $\ga_m$ to the
orders needed later.
For $Z_g$ we get at the two-loop level
\beq
\label{Zgreconstructed}
Z_g=1+\frac{2\be_1 g+\be_2 g^2}{\ep}+\frac{4\be_1^2 g^2}{\ep^2}
+\od(g^3)\,,
\eeq
while the three-loop approximation of $Z_m$ is
\bea
\label{Zmreconstructed}
Z_m
&=&
1+\frac{2\al_1 g+\al_2 g^2+\frac{2}{3}\al_3 g^3}{\ep}
+\frac{2\al_1(\al_1+\be_1)g^2
+2(\al_1\al_2+\frac{2}{3}\al_1\be_2+\frac{2}{3}\al_2\be_1)g^3}{\ep^2}
\nn\\
&&
+\frac{\frac{4}{3}\al_1(\al_1+\be_1)(\al_1+2\be_1)g^3}{\ep^3}
+\od(g^4)\,.
\eea

\section{Absorption of One-Loop Mass Correction into Bare Mass}
%==============================================================
\label{oneloopresummation}

\begin{table}[t]
\begin{center}
\begin{tabular}{c|cccc}
order
&
\multicolumn{4}{c}{diagrams and symmetry factors}
\\
\hline
\rule[-2pt]{0pt}{10pt}
% \begin{tabular}{c}$0$ loops\\$g^{-1}$\end{tabular}&
$\!\!${\footnotesize 0 loops}, $g^{-1}\!\!$&
$1$
\bpi(10,12)
\put(5,3){\circle*{4}}
\epi
\\
\hline
\rule[-10pt]{0pt}{26pt}
$\!\!${\footnotesize 1 loop}, $g^0\!\!$&
\bpi(26,12)
\put(13,3){\circle{16}}
\epi
\\
\hline
\rule[-10pt]{0pt}{26pt}
$\!\!${\footnotesize 2 loops}, $g^1\!\!$&
$\frac{1}{8}$
\bpi(42,12)
\put(13,3){\circle{16}}
\put(29,3){\circle{16}}
\put(21,3){\circle*{4}}
\epi
\\
\hline
\rule[-14pt]{0pt}{34pt}
$\!\!${\footnotesize 3 loops}, $g^2\!\!$&
$\frac{1}{16}$
\bpi(58,12)
\put(13,3){\circle{16}}
\put(29,3){\circle{16}}
\put(45,3){\circle{16}}
\put(21,3){\circle*{4}}
\put(37,3){\circle*{4}}
\epi
&
$\frac{1}{48}$
\bpi(34,12)
\put(17,3){\circle{24}}
\put(17,3){\oval(24,8)}
\put(5,3){\circle*{4}}
\put(29,3){\circle*{4}}
\epi
\\
\hline
\rule[-18pt]{0pt}{54pt}
$\!\!${\footnotesize 4 loops}, $g^3\!\!$&
$\frac{1}{32}$
\bpi(74,12)
\put(13,3){\circle{16}}
\put(29,3){\circle{16}}
\put(45,3){\circle{16}}
\put(61,3){\circle{16}}
\put(21,3){\circle*{4}}
\put(37,3){\circle*{4}}
\put(53,3){\circle*{4}}
\epi
&
$\frac{1}{48}$
\bpi(53.7,12)
\put(26.85,3){\circle{16}}
\put(26.85,19){\circle{16}}
\put(13,-5){\circle{16}}
\put(40.7,-5){\circle{16}}
\put(26.85,11){\circle*{4}}
\put(19.95,-1){\circle*{4}}
\put(33.75,-1){\circle*{4}}
\epi
&
$\frac{1}{24}$
\bpi(34,12)
\put(17,3){\circle{24}}
\put(17,3){\oval(24,8)}
\put(17,23){\circle{16}}
\put(5,3){\circle*{4}}
\put(29,3){\circle*{4}}
\put(17,15){\circle*{4}}
\epi
&
$\frac{1}{48}$
\bpi(34,12)
\put(17,3){\circle{24}}
\put(6.6,9){\line(1,0){20.8}}
\put(6.6,9){\line(3,-5){10.4}}
\put(27.4,9){\line(-3,-5){10.4}}
\put(6.6,9){\circle*{4}}
\put(27.4,9){\circle*{4}}
\put(17,-9){\circle*{4}}
\epi
\end{tabular}
\end{center}
\caption{\protect\label{allgraphs}
         Vacuum graphs up to four loops and their symmetry factors.
         In the equations in the text the symmetry factor is
         considered part of each respective diagram.
         The one-loop graph cannot be constructed by the Feynman
         rules and has to be dealt with separately.
         Therefore it does not carry a symmetry factor.}
\end{table}
Table \ref{allgraphs} shows all vacuum graphs up to four loops,
i.e., to order $g^3$, together with their symmetry factors.
To reduce the number of diagrams to be considered, we will now absorb
the one-loop mass correction into a modified bare mass term in the
Lagrangian.
This will get rid of all diagrams carrying a one-loop correction with
the exception of the two-loop diagram.

Suppose we added a term $\frac{1}{2}\de m_B^2\phi_B^2$ with
$\de m_B^2=\od(g)$ to the free part of our Lagrangian (\ref{Lb})
and subtracted it again in the interaction part.
If then we compute all diagrams to a given order in interaction
vertices and, at the end, reexpand in $g$ to that order, we will
get the same result as if we never made that manipulation.
The changes in the Feynman rules are:
\begin{itemize}
\item Replace $m_B^2$ by $\mb\equiv m_B^2+\de m_B^2$ in the
  propagator.
\item Introduce an additional interaction vertex
 $\bpi(50,12)
  \put(5,0){$a$}
  \put(39,0){$b$}
  \put(15,3){\line(1,0){20}}
  \put(25,3){\circle*{4}}
  \epi
  =Z_\phi\de m_B^2\de_{ab}$\,.
\end{itemize}
\begin{table}[t]
\begin{center}
\begin{tabular}{c|ccc}
order in $g$
&
\multicolumn{3}{c}{diagrams and symmetry factors}
\\
\hline
\rule[-10pt]{0pt}{26pt}
$g^1$&
$\frac{1}{2}$
\bpi(26,12)
\put(13,3){\circle{16}}
\put(21,3){\circle*{4}}
\epi
\\
\hline
\rule[-10pt]{0pt}{26pt}
$g^2$&
$\frac{1}{4}$
\bpi(26,12)
\put(13,3){\circle{16}}
\put(5,3){\circle*{4}}
\put(21,3){\circle*{4}}
\epi
&
$\frac{1}{4}$
\bpi(42,12)
\put(13,3){\circle{16}}
\put(29,3){\circle{16}}
\put(21,3){\circle*{4}}
\put(37,3){\circle*{4}}
\epi
\\
\hline
\rule[-10pt]{0pt}{26pt}
$g^3$&
$\frac{1}{6}$
\bpi(26,12)
\put(13,3){\circle{16}}
\put(13,11){\circle*{4}}
\put(6.1,-1){\circle*{4}}
\put(19.9,-1){\circle*{4}}
\epi
&
$\frac{1}{8}$
\bpi(42,12)
\put(13,3){\circle{16}}
\put(29,3){\circle{16}}
\put(5,3){\circle*{4}}
\put(21,3){\circle*{4}}
\put(37,3){\circle*{4}}
\epi
&
$\frac{1}{4}$
\bpi(42,12)
\put(13,3){\circle{16}}
\put(29,3){\circle{16}}
\put(21,3){\circle*{4}}
\put(33,-3.9){\circle*{4}}
\put(33,9.9){\circle*{4}}
\epi
\\
\rule[-14pt]{0pt}{34pt}
&
$\frac{1}{8}$
\bpi(58,12)
\put(13,3){\circle{16}}
\put(29,3){\circle{16}}
\put(45,3){\circle{16}}
\put(21,3){\circle*{4}}
\put(29,11){\circle*{4}}
\put(37,3){\circle*{4}}
\epi
&
$\frac{1}{8}$
\bpi(58,12)
\put(13,3){\circle{16}}
\put(29,3){\circle{16}}
\put(45,3){\circle{16}}
\put(21,3){\circle*{4}}
\put(37,3){\circle*{4}}
\put(53,3){\circle*{4}}
\epi
&
$\frac{1}{12}$
\bpi(34,12)
\put(17,3){\circle{24}}
\put(17,3){\oval(24,8)}
\put(5,3){\circle*{4}}
\put(17,15){\circle*{4}}
\put(29,3){\circle*{4}}
\epi
\\
\end{tabular}
\end{center}
\caption{\protect\label{additionalgraphs}
         Additional vacuum graphs up to order $g^3$ and their
         symmetry factors as introduced by a quadratic interaction
         vertex of order $g$.}
\end{table}
In Table \ref{additionalgraphs} we list the additional graphs up
to order $g^3$ introduced by this resummation together with their
symmetry factors.

Now choose $\de m_B^2$ such that
\beq
\label{selfenergyabsorption}
\bpi(50,12)
\put(5,0){$a$}
\put(39,0){$b$}
\put(15,3){\line(1,0){20}}
\put(25,3){\circle*{4}}
\epi
+
\rule[-8pt]{0pt}{19pt}
\bpi(60,12)
\put(5,-8){$a$}
\put(49,-8){$b$}
\put(30,3){\circle{16}}
\put(15,-5){\line(1,0){30}}
\put(30,-5){\circle*{4}}
\epi
=0\,.
\eeq
Then $\de m_B^2=\od(g)$ and thus we can carry out the resummation
program of the last paragraph.
Notice however, that when summing the diagrams of Tables
\ref{allgraphs} and \ref{additionalgraphs} (keeping in mind that
the symmetry factors are considered part of the diagrams here and
are not multiplying the diagrams), most of them cancel through
relation (\ref{selfenergyabsorption}).
\begin{table}[t]
\begin{center}
\begin{tabular}{c|c|c}
\begin{tabular}{c}number \\of loops\end{tabular}
&
order in $g$
&
\begin{tabular}{c}remaining diagrams and\\
revised symmetry factors\end{tabular}
\\
\hline
\rule[-2pt]{0pt}{10pt}
0&$g^{-1}\!\!\!\!$&
$1$
\bpi(10,12)
\put(5,3){\circle*{4}}
\epi
\\
\hline
\rule[-10pt]{0pt}{26pt}
1&$g^0$&
\bpi(26,12)
\put(13,3){\circle{16}}
\epi
\\
\hline
\rule[-10pt]{0pt}{26pt}
2&$g^1$&
$-\frac{1}{8}$
\bpi(42,12)
\put(13,3){\circle{16}}
\put(29,3){\circle{16}}
\put(21,3){\circle*{4}}
\epi
\\
\hline
\rule[-14pt]{0pt}{34pt}
3&$g^2$&
$\frac{1}{48}$
\bpi(34,12)
\put(17,3){\circle{24}}
\put(17,3){\oval(24,8)}
\put(5,3){\circle*{4}}
\put(29,3){\circle*{4}}
\epi
\\
\hline
\rule[-14pt]{0pt}{34pt}
4&$g^3$&
$\frac{1}{48}$
\bpi(34,12)
\put(17,3){\circle{24}}
\put(6.6,9){\line(1,0){20.8}}
\put(6.6,9){\line(3,-5){10.4}}
\put(27.4,9){\line(-3,-5){10.4}}
\put(6.6,9){\circle*{4}}
\put(27.4,9){\circle*{4}}
\put(17,-9){\circle*{4}}
\epi
\end{tabular}
\end{center}
\caption{\protect\label{remaininggraphs}
         Remaining diagrams after resummation of the quadratic
         part of the Lagrangian.}
\end{table}
The only remaining diagrams to order $g^3$ are listed in Table
\ref{remaininggraphs}.
We thus have succeeded in eliminating all diagrams with one-loop mass
corrections with the exception of the two-loop diagram for which the
symmetry factor does not work out, since each of the two bubbles can
act as a correction to the other one.

Next we have to solve (\ref{selfenergyabsorption}) for $\mb$.
With
\beq
\bpi(50,12)
\put(5,0){$a$}
\put(39,0){$b$}
\put(15,3){\line(1,0){20}}
\put(25,3){\circle*{4}}
\epi
=\de_{ab}Z_\phi\de m_B^2
=\de_{ab}Z_\phi(\mb-m_B^2)
\eeq
and
\bea
\rule[-8pt]{0pt}{19pt}
\bpi(60,12)
\put(5,-8){$a$}
\put(49,-8){$b$}
\put(30,3){\circle{16}}
\put(15,-5){\line(1,0){30}}
\put(30,-5){\circle*{4}}
\epi
&=&
\frac{1}{2}\left(-Z_\phi^2\frac{(4\pi)^2\mu^{-\ep}g_B}{3}\right)
[\de_{ab}\de_{cc}+2\de_{ac}\de_{bc}]
\int_p\frac{Z_\phi^{-1}}{p^2+\mb}
\nn\\
&=&
-\frac{\de_{ab}(N+2)(4\pi)^2I_{1A}}{6}Z_\phi Z_g g
\left(\frac{\mb}{m^2}\right)^{1-\frac{\ep}{2}}
\eea
with $I_{1A}$ from (\ref{i1a}), we get
\beq
\label{gapequation}
\mb=m_B^2+\frac{(N+2)(4\pi)^2I_{1A}}{6}Z_g g
\left(\frac{\mb}{m^2}\right)^{1-\frac{\ep}{2}}
\eeq
as the defining equation for $\mb$.
This cannot be solved explicitly.
However, we are only interested in the first few terms of a
power series of $\mb$ in $g$.
Define $a_l$ and $\bar{a}_l$ by
\bea
m_B^2 &=& Z_m m^2 = m^2\left(1+\sum_{l=1}^\infty a_l g^l\right)\,,
\\
\label{mbar}
\mb &=& Z_{\bar{m}} m^2 =
m^2\left(1+\sum_{l=1}^\infty \bar{a}_l g^l\right)\,.
\eea
The $a_l$ can be read off from (\ref{Zmreconstructed}) to be
\bea
\label{a}
\ba{ll}
a_1=&\ds\!\!\frac{2\al_1}{\ep}\,,
\vs\\
a_2=&\ds\!\!\frac{\al_2}{\ep}+\frac{2\al_1(\al_1+\be_1)}{\ep^2}\,,
\vs\\
\ds
a_3=&\ds\!\!\frac{\frac{2}{3}\al_3}{\ep}
+\frac{+2(\al_1\al_2+\frac{2}{3}\al_1\be_2
+\frac{2}{3}\al_2\be_1)}{\ep^2}
+\frac{\frac{4}{3}\al_1(\al_1+\be_1)(\al_1+2\be_1)}{\ep^3}\,.
\ea
\eea
Further, define $b_l$ by
\beq
\frac{(N+2)(4\pi)^2}{6}\frac{I_{1A}}{m^2}Z_g g
=\sum_{l=1}^\infty b_l g^l\,,
\eeq
such that the $b_l$ are given by
\bea
\ba{ll}
b_1 &\ds\!\!= \frac{(N+2)(4\pi)^2}{6}\frac{I_{1A}}{m^2}\,,
\vs\\
b_l &\ds\!\!= \frac{(N+2)(4\pi)^2}{6}\frac{I_{1A}}{m^2}
\sum_{k=1}^{l-1}\frac{Z_{k,l-1}^g}{\ep^k}
\he
l>1\,.
\ea
\eea
That is, with the help of (\ref{zs}) and (\ref{Zgreconstructed})
we can write
\beq
\label{b}
\ba{ll}
b_1 &\ds\!\!= \frac{(N+2)(4\pi)^2}{6}\frac{I_{1A}}{m^2}\,,
\vs\\
b_2 &\ds\!\!= \frac{(N+2)(4\pi)^2}{6}\frac{I_{1A}}{m^2}
\frac{2\be_1}{\ep}\,,
\vs\\
b_3 &\ds\!\!= \frac{(N+2)(4\pi)^2}{6}\frac{I_{1A}}{m^2}
\left(\frac{\be_2}{\ep}+\frac{4\be_1^2}{\ep^2}\right)\,.
\ea
\eeq
Now we can restate (\ref{gapequation}) as
\beq
\sum_{l=1}^\infty \bar{a}_l g^l
=\sum_{l=1}^\infty a_l g^l+\sum_{l=1}^\infty b_l g^l
\left(1+\sum_{l=1}^\infty \bar{a}_l g^l\right)^{1-\frac{\ep}{2}}\,.
\eeq
Expanding in powers of $g$ and comparing coefficients of powers of
$g$, we get the relations
\beq
\label{abar}
\ba{ll}
\bar{a}_1
&\ds\!\!=
a_1+b_1\,,
\vs\\
\bar{a}_2
&\ds\!\!=
a_2+b_2+\left(1-\frac{\ep}{2}\right)\bar{a}_1 b_1\,,
\vs\\
\bar{a}_3
&\ds\!\!=
a_3+b_3
+\left(1-\frac{\ep}{2}\right)\left[\bar{a}_1 b_2+\bar{a}_2 b_1
+\left(-\frac{\ep}{2}\right)\frac{\bar{a}_1^2}{2}b_1\right]\,,
\ea
\eeq
which recursively define the three coefficients needed for a
four-loop computation of $\be_v$.

Finally, the effective Feynman rules to be used for our vacuum
diagrams are
\beq
\label{modified feynman rules}
\!\!
\ba{ll}
{\rm \underline{propagator:}}
\\
\hh
\bpi(50,12)
\put(5,0){$a$}
\put(39,0){$b$}
\put(15,3){\line(1,0){20}}
\epi
&\ds
=\frac{\de_{ab}Z_\phi^{-1}}{p^2+\mb}
\vs\\
{\rm \underline{vertices:}}
\\
\hh
\bpi(50,12)
\put(25,3){\circle*{4}}
\epi
&\ds
= -\frac{m_B^4 h_B}{(4\pi)^2\mu^{-\ep}g_B}
\vs\\
\hh
\rule[-10pt]{0pt}{28pt}
\bpi(50,12)
\put(5,10){$a$}
\put(39,10){$b$}
\put(5,-10){$c$}
\put(39,-10){$d$}
\put(15,-7){\line(1,1){20}}
\put(15,13){\line(1,-1){20}}
\put(25,3){\circle*{4}}
\epi
\hh
&\ds
=-[\de_{ab}\de_{cd}+\de_{ac}\de_{bd}+\de_{ad}\de_{bc}]
Z_\phi^2\,\frac{(4\pi)^2\mu^{-\ep}g_B}{3}
\\
\ea
\eeq
with $\mb$ given at the three-loop level by (\ref{mbar}), (\ref{a}),
(\ref{b}), (\ref{abar}) and the integral $I_{1A}$ from (\ref{i1a}).
Note that the mass has changed only in the propagator, not in the
zero-loop diagram represented by the dot.
The loop momentum integration measure is again given by
(\ref{integration measure}).
Only the diagrams in Table \ref{remaininggraphs} are to be
calculated.
The new symmetry factors are stated there, too.
As a general rule, only diagrams without a one-loop mass correction
have to be computed with the exception of the two-loop diagram.
The two-loop diagram changes sign, while all the other diagrams
to be computed have their standard symmetry factor.

\section{$\be_v$ and $Z_v$ to Four Loops}
%========================================
\label{fourloop}

In order to achieve a four-loop computation of $\be_v$, we have to
keep all terms in zero to four loops up to order $g^3$.
Since we are interested only in the divergent part of diagrams, we
will disregard terms of order $\ep^0$.

Our strategy will be to use the bare coupling and modified bare mass
expressed in terms of the coefficients $\be_k$ and $\al_k$ and then
to postulate the appropriate cancellation of subdivergencies by
counterterms, which in practice means the demand that $Z_v$ does not
contain any logarithms of the renormalized mass.
When determining $\be_v$ and $Z_v$ to $k$ loops by using this
procedure, we will get as a by-product $\be_{g}$ to $k-2$ loops
(since $g$ effectively has its first vacuum loop graph appearance at
two loops) and $\ga_{m}$ to $k-1$ loops (since for $m^2$ this
appearance is at one loop).

\subsection{One Loop}
%--------------------

Using the modified Feynman rules (\ref{modified feynman rules}),
the one-loop diagram is evaluated as
\beq
\rule[-5pt]{0pt}{16pt}
\bpi(26,12)
\put(13,3){\circle{16}}
\epi
=
\frac{\de_{aa}}{2}\int_p\ln\frac{Z_\phi^{-1}}{p^2+\mb}
=
-\frac{N}{2}Z_{\bar{m}}^{2-\frac{\ep}{2}}I_1\,,
\eeq
where $I_1\equiv\int_p\ln(p^2+m^2)$.
With $Z_{\bar{m}}$ given by (\ref{mbar}), (\ref{a}), (\ref{b}) and
(\ref{abar}), and $I_1$ from (\ref{i1}), one gets
\beq
\bpi(14,12)
\put(7,3){\circle*{4}}
\epi
+
\rule[-5pt]{0pt}{16pt}
\bpi(26,12)
\put(13,3){\circle{16}}
\epi
=
-\frac{m^4}{(4\pi)^2g}\left[h+\left(Z_{11}^v
-\frac{N}{2}\right)\frac{g}{\ep}+\od(g^2,\ep^0)\right]\,.
\eeq
Demanding this to be finite as $\ep\rightarrow 0$ gives
\beq
\label{z11}
Z_{11}^v=\frac{N}{2}\,.
\eeq

\subsection{Two Loops}
%---------------------

Using again the modified Feynman rules
(\ref{modified feynman rules}), the two-loop diagram is evaluated
as (remember that the two-loop diagram now enters with the opposite
sign than usual)
\beq
\rule[-5pt]{0pt}{16pt}
\bpi(42,12)
\put(13,3){\circle{16}}
\put(29,3){\circle{16}}
\put(21,3){\circle*{4}}
\epi
=\frac{N(N+2)g}{24}Z_g Z_{\bar{m}}^{2-\ep}I_{1A}^2
\eeq
with $I_{1A}$ defined in (\ref{i1a}).
Plugging in $Z_{\bar{m}}$ again, using $Z_g$ from
(\ref{Zgreconstructed}) and $I_{1A}$ from (\ref{i1a}) and observing
(\ref{z11}), one gets
\bea
\rule[-10pt]{0pt}{26pt}
\bpi(14,12)
\put(7,3){\circle*{4}}
\epi
+
\bpi(26,12)
\put(13,3){\circle{16}}
\epi
+
\bpi(42,12)
\put(13,3){\circle{16}}
\put(29,3){\circle{16}}
\put(21,3){\circle*{4}}
\epi
&=&
-\frac{m^4}{(4\pi)^2g}\left\{h+
\left[Z_{12}^v+N\left(\al_1-\frac{N+2}{6}\right)
\left(\lnma-1\right)\right]\frac{g^2}{\ep}\right.
\nn\\
&&
\he\left.+\left[Z_{22}^v-2N\left(\al_1-\frac{N+2}{12}\right)\right]
\frac{g^2}{\ep^2}+\od(g^3,\ep^0)\right\}\,.
\eea
Demanding this to be finite as $\ep\rightarrow 0$ and that the
$Z_{kl}^v$ contain no logarithms gives
\beq
\label{alpha1}
\al_1=\frac{N+2}{6}
\eeq
and
\beq
\label{z12z22}
Z_{12}^v=0\,,\he Z_{22}^v=\frac{N(N+2)}{6}\,.
\eeq

\subsection{Three Loops}
%-----------------------

The three-loop diagram is evaluated as
\beq
\rule[-14pt]{0pt}{34pt}
\bpi(34,12)
\put(17,3){\circle{24}}
\put(17,3){\oval(24,8)}
\put(5,3){\circle*{4}}
\put(29,3){\circle*{4}}
\epi
=\frac{N(N+2)g^2}{144}
Z_g^2 Z_{\bar{m}}^{2-\frac{3}{2}\ep}I_2^{cc}\,,
\eeq
where $I_2^{cc}$ belongs to the class of circle-chain integrals
defined in (\ref{irc}).

Plugging in $Z_{\bar{m}}$ and $Z_g$ and with $I_2^{cc}$ from
(\ref{i2rcres}) and making use of (\ref{z11}), (\ref{alpha1}) and
(\ref{z12z22}), one gets
\bea
\lefteqn{
\rule[-14pt]{0pt}{34pt}
\bpi(14,12)
\put(7,3){\circle*{4}}
\epi
+
\bpi(26,12)
\put(13,3){\circle{16}}
\epi
+
\bpi(42,12)
\put(13,3){\circle{16}}
\put(29,3){\circle{16}}
\put(21,3){\circle*{4}}
\epi
+
\bpi(34,12)
\put(17,3){\circle{24}}
\put(17,3){\oval(24,8)}
\put(5,3){\circle*{4}}
\put(29,3){\circle*{4}}
\epi
}
\nn\\
&=&
\left.-\frac{m^4}{(4\pi)^2g}\right\{h
\nn\\
&&
\he+\left[Z_{13}^v
+\frac{N(N+2)}{8}\right.\left(\be_1-\frac{N+8}{6}\right)
\left(\lnma-1\right)^2
+\frac{N}{2}\left(\al_2+\frac{5(N+2)}{36}\right)\left(\lnma-1\right)
\nn\\
&&
\he\hf\hh\left.
+\frac{N(N+2)}{24}\left[[1+\ze(2)]\left(\be_1-\frac{N+8}{6}\right)
-\frac{1}{6}\right]
\right]\frac{g^3}{\ep}
\nn\\
&&
\he+\left[Z_{23}^v-\frac{N(N+2)}{6}\left(\be_1-\frac{N+8}{6}\right)
\left(\lnma-1\right)-N\left(\al_2+\frac{5(N+2)}{108}\right)\right]
\frac{g^3}{\ep^2}
\nn\\
&&
\he\left.+\left[Z_{33}^v-\frac{N(N+2)(N+4)}{18}\right]
\frac{g^3}{\ep^3}+\od(g^4,\ep^0)\right\}\,.
\eea
Demanding this to be finite as $\ep\rightarrow 0$ and that the
$Z_{kl}^v$ contain no logarithms gives
\beq
\label{beta1alpha2}
\be_1=\frac{N+8}{6}\,,\he\al_2=-\frac{5(N+2)}{36}\,,
\eeq
and
\beq
\label{z13z23z33}
\he Z_{13}^v=\frac{N(N+2)}{144}\,,
\he Z_{23}^v=-\frac{5N(N+2)}{54}\,,
\he Z_{33}^v=\frac{N(N+2)(N+4)}{18}\,.
\eeq

\subsection{Four Loops}
%----------------------

The four-loop diagram is evaluated as
\beq
\bpi(34,12)
\put(17,3){\circle{24}}
\put(6.6,9){\line(1,0){20.8}}
\put(6.6,9){\line(3,-5){10.4}}
\put(27.4,9){\line(-3,-5){10.4}}
\put(6.6,9){\circle*{4}}
\put(27.4,9){\circle*{4}}
\put(17,-9){\circle*{4}}
\epi
=-\frac{N(N+2)(N+8)g^3}{1296}
Z_g^3 Z_{\bar{m}}^{2-2\ep}I_3^{cc}\,,
\eeq
where $I_3^{cc}$ also belongs to the class of circle-chain integrals
defined in (\ref{irc}).
Plugging in $Z_{\bar{m}}$ and $Z_g$ and with $I_3^{cc}$ from
(\ref{i3rc}) and making use of (\ref{z11}), (\ref{alpha1}),
(\ref{z12z22}), (\ref{beta1alpha2}) and (\ref{z13z23z33}), one gets
\bea
\lefteqn{
\rule[-14pt]{0pt}{34pt}
\bpi(14,12)
\put(7,3){\circle*{4}}
\epi
+
\bpi(26,12)
\put(13,3){\circle{16}}
\epi
+
\bpi(42,12)
\put(13,3){\circle{16}}
\put(29,3){\circle{16}}
\put(21,3){\circle*{4}}
\epi
+
\bpi(34,12)
\put(17,3){\circle{24}}
\put(17,3){\oval(24,8)}
\put(5,3){\circle*{4}}
\put(29,3){\circle*{4}}
\epi
+
\bpi(34,12)
\put(17,3){\circle{24}}
\put(6.6,9){\line(1,0){20.8}}
\put(6.6,9){\line(3,-5){10.4}}
\put(27.4,9){\line(-3,-5){10.4}}
\put(6.6,9){\circle*{4}}
\put(27.4,9){\circle*{4}}
\put(17,-9){\circle*{4}}
\epi
}
\nn\\
&=&
\left.-\frac{m^4}{(4\pi)^2g}\right\{h
\nn\\
&&
\he+\left[Z_{14}^v
+\frac{N(N+2)}{18}\left(\be_2+\frac{3N+14}{6}\right)
\left(\lnma-1\right)^2\right.
\nn\\
&&
\he\hf
+\frac{N}{3}\left(\al_3-\frac{(N+2)(5N+37)}{72}\right)
\left(\lnma-1\right)
\nn\\
&&
\he\hf\left.
+\frac{N(N+2)}{72}\left(\frac{43N}{36}+\frac{71}{9}+\be_2
+\left(\be_2+\frac{3N+14}{6}\right)\ze(2)
-\frac{N+8}{3}\ze(3)\right)
\right]\frac{g^4}{\ep}
\nn\\
&&
\he+\left[Z_{24}^v
-\frac{N(N+2)}{18}\left(\be_2+\frac{3N+14}{6}\right)
\left(\lnma-1\right)
-\frac{2N}{3}\left(\al_3-\frac{(N+2)(N+8)}{72}\right)
\right]
\frac{g^4}{\ep^2}
\nn\\
&&
\he+\left[Z_{34}^v
-\frac{N(N+2)}{18}\left(\be_2-\frac{11N+43}{9}\right)
\right]
\frac{g^4}{\ep^3}
\nn\\
&&
\he\left.+\left[Z_{44}^v-\frac{N(N+2)(N+4)(N+5)}{54}\right]
\frac{g^4}{\ep^4}+\od(g^5,\ep^0)\right\}\,.
\eea
Demanding this to be finite as $\ep\rightarrow 0$ and that the
$Z_{kl}^v$ contain no logarithms gives
\beq
\label{beta2alpha3}
\be_2=-\frac{3N+14}{6}\,,\he\al_3=\frac{(N+2)(5N+37)}{72}\,,
\eeq
and
\bea
\label{z14z24z34z44}
\ba{ll}
\ds Z_{14}^v=\frac{N(N+2)(N+8)(12\ze(3)-25)}{2592}\,,\he
&
\ds Z_{24}^v=\frac{N(N+2)(4N+29)}{108}\,,
\vs\\
\ds Z_{34}^v=-\frac{N(N+2)(31N+128)}{324}\,,
&
\ds Z_{44}^v=\frac{N(N+2)(N+4)(N+5)}{54}\,.
\ea
\eea

\subsection{Check of Recursion Relations for the $Z_{kl}^v$}
%-----------------------------------------------------------

In this section we check the recursion relations between the
$Z_{kl}^v$ we have computed.
Putting (\ref{zs}) and (\ref{zrecursion}) together and separating
into powers of $g$ we get
\beq
Z_{k+1,n+1}^v=\frac{1}{n+1}
\sum_{l=1}^{n-k+1}l[2Z_{1l}^m+(n-l)Z_{1l}^g]Z_{k,n-l+1}^v\,,
\he 1\leq k\leq n\,.
\eeq
Note that to verify the recursion relations for the $(n+1)$-loop
order coefficients $Z_{k+1,n+1}^v$ for all $k$ with $1\leq k\leq n$,
we need $Z_{m,1}$ to $n$-loop order and, because of the $(n-l)$
factor, $Z_{g,1}$ only to $(n-1)$-loop order.

The relevant relations are
\beq
\ba{l}
Z_{22}^v
=
\frac{1}{2}2Z_{11}^m Z_{11}^v\,,
\vs\\
Z_{23}^v
=
\frac{1}{3}[(2Z_{11}^m+Z_{11}^g)Z_{12}^v+2Z_{12}^mZ_{11}^v]\,,
\vs\\
Z_{33}^v
=
\frac{1}{3}(2Z_{11}^m+Z_{11}^g)Z_{22}^v\,,
\vs\\
Z_{24}^v
=
\frac{1}{4}[(2Z_{11}^m+2Z_{11}^g)Z_{13}^v
+2(2Z_{12}^m+Z_{12}^g)Z_{12}^v+3(2Z_{13}^m)Z_{11}^v]\,,
\vs\\
Z_{34}^v
=
\frac{1}{4}[(2Z_{11}^m+2Z_{11}^g)Z_{23}^v
+2(2Z_{12}^m+Z_{12}^g)Z_{22}^v]\,,
\vs\\
Z_{44}^v
=
\frac{1}{4}(2Z_{11}^m+2Z_{11}^g)Z_{33}^v\,.
\ea
\eeq
The $Z_{kl}^v$ involved are given in (\ref{z11}), (\ref{z12z22}),
(\ref{z13z23z33}) and (\ref{z14z24z34z44}).
The necessary $Z_{kl}^g$ and $Z_{kl}^m$ can be constructed with the
help of (\ref{zs}), (\ref{Zgreconstructed}) and
(\ref{Zmreconstructed}), using the $\be_k$ and $\al_k$ from
(\ref{alpha1}), (\ref{beta1alpha2}) and (\ref{beta2alpha3}).

It is straightforward to check that all of the above recursion
relations hold.
Also, the values found for $\be_1$, $\be_2$, $\al_1$, $\al_2$ and
$\al_3$ coincide with those in the literature, see e.g.\ \cite{Kl..}.

\subsection{$\be_v$ to Four Loops}
%---------------------------------

Constructing $Z_{v,1}$ to four loops from (\ref{z11}),
(\ref{z12z22}), (\ref{z13z23z33}) and (\ref{z14z24z34z44}) and
using (\ref{begares}), we get our final result,
\beq
\be_v=\frac{N}{4}g+\frac{N(N+2)}{96}g^3
+\frac{N(N+2)(N+8)[12\ze(3)-25]}{1296}g^4+\od(g^5)\,.
\eeq

It would be worthwhile to continue to higher loops to be able to
make meaningful comparisons of divergencies of vacuum diagrams
with invariants of knot theory and to try to derive $\be_g$ and
$\ga_m$ to higher loops with this method as well.
Also, it would be interesting to investigate possible connections
of $\be_v$ via the $\ep$ expansion with the critical theory in
three dimensions.

\section*{Acknowledgements}
%==========================

I would like to thank T.~Clark, S.~Love and S.~Khlebnikov for
useful discussions and H.~Kleinert for beneficial communications.
This work was supported by the U.S.\ Department of Energy,
contract No.\ DE-FG02-91ER40681 (Task B).

\appendix
%========
\renewcommand{\thesection}{Appendix \Alph{section}}

\section{$I_1$, $I_{1A}$ and $I_{1B}$}
%=====================================

Using standard methods, $I_1$, $I_{1A}$ and $I_{1B}$ are evaluated as
\bea
\label{i1}
I_1
&\equiv&
\int_p\ln(p^2+m^2)
=-\frac{m^4}{(4\pi)^2}
\left(\frac{m^2}{4\pi\mu^2}\right)^{-\frac{\ep}{2}}
\Ga(\ts\frac{\ep}{2}-2)
\,,
\\
\label{i1a}
I_{1A}
&\equiv&
\int_p\frac{1}{p^2+m^2}
=\frac{m^2}{(4\pi)^2}
\left(\frac{m^2}{4\pi\mu^2}\right)^{-\frac{\ep}{2}}
\Ga(\ts\frac{\ep}{2}-1)
\,,
\\
\label{i1b}
I_{1B}
&\equiv&
\int_p\frac{1}{(p^2+m^2)^2}
=\frac{1}{(4\pi)^2}
\left(\frac{m^2}{4\pi\mu^2}\right)^{-\frac{\ep}{2}}
\Ga(\ts\frac{\ep}{2})
\,.
\eea

\section{$I_2^{cc}$}
%===================

Since
\beq
I_2^{cc}
\equiv\int_{pqr}\frac{1}{[(p+q+r)^2+m^2](p^2+m^2)(q^2+m^2)(r^2+m^2)}
\propto(m^2)^{\frac{3}{2}d-4}
\eeq
we can write
\beq
I_2^{cc}
=\frac{1}{\frac{3}{2}d-4}m^2\frac{\p}{\p m^2}I_2^{cc}
=-\frac{8m^2}{3d-8}\int_{pqr}
\frac{1}{[(p+q+r)^2+m^2]^2(p^2+m^2)(q^2+m^2)(r^2+m^2)}\,.
\eeq
Using some simple algebra on the integrand, this can be rewritten as
\beq
I_2^{cc}=I_{2a}^{cc}+I_{2b}^{cc}+I_{2c}^{cc}+I_{2d}^{cc}
\eeq
with
\bea
\ba{l}
\ds I_{2a}^{cc}
=\frac{16m^2}{4-3\ep}\int_{pqr}
\frac{1}{[(p+q+r)^2+m^2]^2p^2q^2r^2}\,,
\vs\\
\ds I_{2b}^{cc}
=-\frac{24m^2}{4-3\ep}\int_{pqr}
\frac{1}{[(p+q+r)^2+m^2]^2p^2q^2(r^2+m^2)}\,,
\vs\\
\ds I_{2c}^{cc}
=-\frac{24m^6}{4-3\ep}\int_{pqr}
\frac{1}{[(p+q+r)^2+m^2]^2(p^2+m^2)p^2(q^2+m^2)q^2r^2}\,,
\vs\\
\ds I_{2d}^{cc}
=\frac{8m^8}{4-3\ep}\int_{pqr}
\frac{1}{[(p+q+r)^2+m^2]^2(p^2+m^2)p^2(q^2+m^2)q^2(r^2+m^2)r^2}\,.
\ea
\eea
$I_{2c}^{cc}$ and $I_{2d}^{cc}$ are UV finite in four dimensions.
The evaluation of $I_{2a}^{cc}$ and $I_{2b}^{cc}$ is
straightforward using standard methods.
The results are
\beq
I_{2a}^{cc}
=\frac{2m^4}{(4\pi)^6}\left[\frac{4}{3\ep^2}+\frac{1}{\ep}\left(
5-2\lnma\right)\right]+\od(\ep^0)
\eeq
and
\beq
I_{2b}^{cc}
=\frac{m^4}{(4\pi)^6}
\left[\frac{16}{\ep^3}+\frac{1}{\ep^2}\left(28-24\lnma\right)
+\frac{1}{\ep}\left(25-42\lnma+18\lnmb+6\ze(2)\right)\right]
+\od(\ep^0)\,.
\eeq
Therefore,
\beq
\label{i2rcres}
I_2^{cc}
=\frac{m^4}{(4\pi)^6}
\left[\frac{16}{\ep^3}
+\frac{1}{\ep^2}\left(\frac{92}{3}-24\lnma\right)
+\frac{1}{\ep}\left(35-46\lnma
+18\lnmb+6\ze(2)\right)\right]
+I_{2,f}^{cc}\,,
\eeq
where $I_{2,f}^{cc}=\od(\ep^0)$.
In order to limit the source of $\pi$'s to phase space factors,
we do not evaluate $\ze(2)=\pi^2/6$ here.

\section{General Circle-Chain Integrals $I_n^{cc}$}
%==================================================

In this section we show how to deal with the circle-chain integrals
defined by
\beq
\label{irc}
I_n^{cc}\equiv\int_k\th(k^2)^n\,,
\he\th(k^2)\equiv\int_p\frac{1}{[(k+p)^2+m^2](p^2+m^2)}\,,
\eeq
which are needed for diagrams of the form
\beq
\rule[-22pt]{0pt}{50pt}
\bpi(50,12)
\put(25,3){\oval(40,40)[b]}
\put(25,3){\oval(40,40)[tl]}
\put(25,21.48){\line(-1,0){7.65}}
\put(17.35,21.48){\line(-1,-1){10.82}}
\put(6.52,10.65){\line(0,-1){15.31}}
\put(6.52,-4.65){\line(1,-1){10.82}}
\put(17.35,-15.48){\line(1,0){15.31}}
\put(32.65,-15.48){\line(1,1){10.82}}
\put(43.48,-4.65){\line(0,1){7.65}}
\put(39.14,17.14){\line(1,0){0.1}}
\put(32.65,21.48){\line(1,0){0.1}}
\put(43.48,10.65){\line(1,0){0.1}}
\put(28.90,22.62){\line(1,0){0.1}}
\put(36.11,19.63){\line(1,0){0.1}}
\put(41.63,14.11){\line(1,0){0.1}}
\put(44.62, 6.90){\line(1,0){0.1}}
\epi\,\,.
\eeq
Note that the two-, three- and four-loop diagrams of Table
\ref{remaininggraphs} are all of this form, and in all higher loop
orders there is one diagram of this form, too.

First, separate $\th(k^2)$ into a divergent part $\th_d$,
independent of $k^2$, and a finite, $k^2$-dependent part $\th_f(k^2)$
according to
\beq
\label{df}
\th(k^2)=\th_d+\th_f(k^2)
\eeq
with
\beq
\th_d=\int_p\frac{1}{(p^2+m^2)^2}=I_{1B}\,,\he
\th_f(k^2)
=\int_p\frac{1}{p^2+m^2}\left(\frac{1}{(k+p)^2+m^2}
-\frac{1}{p^2+m^2}\right)\,,
\eeq
with $I_{1B}$ from (\ref{i1b}).

It is useful to establish the recursion relation
\beq
\label{ircrecursion}
I_n^{cc}
=\int_k\th_f(k^2)^n+\sum_{k=1}^{n-1}
\left(\ba{c}n\\k\ea\right)(-1)^{n-k+1}\th_d^{n-k}I_k^{cc}\,,
\eeq
which follows easily from (\ref{irc}), (\ref{df}) and the fact that
$I_0^{cc}=0$.
For each loop order we will compute the divergent part of
$\int_k\th_f(k^2)^n$.
For $n\geq 2$, we will denote the finite part of $I_n^{cc}$
by $I_{n,f}^{cc}$, such that due to the divergent nature of $\th_d$
we will not determine the divergent part of $I_n^{cc}$ completely.
The remedy of the situation will be the cancellation of the
divergent prefactors of the $I_{n,f}^{cc}$, once the counterterms
are properly taken into account.
For $n=1$, we will also consider the convergent part to the order
needed to avoid writing $I_{1,f}^{cc}$.
This is necessary since we are not considering the one- and two-loop
subdivergencies in a consistent way that would allow us to avoid the
appearing logarithms from the outset.
However, this is no problem, since $I_1^{cc}$ is the square of a
simple one-loop integral and can be computed to arbitrarily high
order in $\ep$.

Now turn to $\th_f(k^2)$. Define
\beq
\de\equiv\frac{4m^2}{k^2+4m^2}
\eeq
and use standard methods to write
\beq
\label{thetafeval}
\th_f(k^2)
=2k_\mu\int_0^1 d\al\,\al
\int_p\frac{2p_\mu+k_\mu}{[p^2+2\al p\cdot k+\al k^2+m^2]^3}
\nn\\
=\frac{4\Ga(\frac{\ep}{2}+1)}
{(4\pi)^2}\left(\frac{m^2}{4\pi\mu^2}\right)^{-\frac{\ep}{2}}\th_\de
\eeq
with
\beq
\th_\de
=(1-\de)\de^\frac{\ep}{2}\int_0^1\frac{d\al\,\al(1-2\al)}
{[4\al(1-\al)+(1-2\al)^2\de]^{1+\frac{\ep}{2}}}
=-\frac{(1-\de)\de^\frac{\ep}{2}}{4}
\int_0^1
\frac{d\be\,\be^\frac{1}{2}}{[1+(\de-1)\be]^{1+\frac{\ep}{2}}}\,,
\eeq
where in the last step we changed variables
according to $\be=(1-2\al)^2$.
Use (\ref{hgex})-(\ref{hgelrep}) to write
\bea
\label{thetadeltaeval}
\th_\de
&=&
-\frac{(1-\de)\de^\frac{\ep}{2}}{6}
{\ts F(1+\frac{\ep}{2},\frac{3}{2};\frac{5}{2};1-\de)}
\nn\\
&=&
-\frac{(1-\de)\de^\frac{\ep}{2}}{6}
{\ts\Ga(\frac{5}{2})}
\left[\frac{\Ga(-\frac{\ep}{2})
{\ts F(1+\frac{\ep}{2},\frac{3}{2};1+\frac{\ep}{2};\de)}}
{\Ga(\frac{3}{2}-\frac{\ep}{2})\Ga(1)}
+\frac{\de^{-\frac{\ep}{2}}\Ga(\frac{\ep}{2})
{\ts F(\frac{3}{2}-\frac{\ep}{2},1;1-\frac{\ep}{2};\de)}}
{\Ga(1+\frac{\ep}{2})\Ga(\frac{3}{2})}
\right]
\nn\\
&=&
-\frac{1-\de}{4}
\left[\frac{\Ga(\frac{3}{2})\Ga(-\frac{\ep}{2})}
{\Ga(\frac{3}{2}-\frac{\ep}{2})}
\de^\frac{\ep}{2}(1-\de)^{-\frac{3}{2}}
+\frac{2}{\ep}{\ts
F(\frac{3}{2}-\frac{\ep}{2},1;1-\frac{\ep}{2};\de)}
\right]
\nn\\
&=&
-\frac{\Ga(\frac{3}{2})\Ga(-\frac{\ep}{2})}
{4\Ga(\frac{3}{2}-\frac{\ep}{2})}
\de^\frac{\ep}{2}(1-\de)^{-\frac{1}{2}}
-\frac{1-\de}{2\ep}
{\ts F(\frac{3}{2}-\frac{\ep}{2},1;1-\frac{\ep}{2};\de)}
\nn\\
&=&
-\frac{\Ga(\frac{3}{2})\Ga(-\frac{\ep}{2})}
{4\Ga(\frac{3}{2}-\frac{\ep}{2})}
\left[{\ts\de^\frac{\ep}{2}+\frac{1}{2}\de^{1+\frac{\ep}{2}}
+\frac{3}{8}\de^{2+\frac{\ep}{2}}}\right]
-\frac{1}{2\ep}
\left[{\ts 1+\frac{1}{2-\ep}\de+\frac{3-\ep}{(2-\ep)(4-\ep)}\de^2}
\right]
+\od(\de^3,\de^{3+\frac{\ep}{2}})\,,
\eea
where $F(\al,\be;\ga;z)$ is Gauss's hypergeometric function,
see \ref{hg}.
Expansion in powers of $\ep$ shows that this expression is indeed
convergent as $\ep\rightarrow 0$:
Despite the appearance of $\Ga(\ts\frac{\ep}{2})$ and
$\ts\frac{2}{\ep}$, no additional UV divergences are introduced and
the only UV divergence in $\int_k\th_f(k^2)^n$ comes from the
$k$ integral.
Note that a spurious IR divergence appeared in the next-to-last line
of (\ref{thetadeltaeval}) as $(1-\de)^{-\frac{1}{2}}$.
However, we know that $\th_\de$ is convergent (in fact: zero) as
$\de\rightarrow 1$ and expanding consistently in $\de$ gets rid
of this intermediate IR divergence.
In other words, there is a cancelling IR divergence in the
hypergeometric function on the same line.

With (\ref{thetafeval}), we get
\bea
\label{thf}
\th_f(k^2)
&=&
\frac{\Ga(\frac{\ep}{2})}{(4\pi)^2}
\left(\frac{m^2}{4\pi\mu^2}\right)^{-\frac{\ep}{2}}
\left\{\frac{\Ga(\frac{3}{2})\Ga(1-\frac{\ep}{2})}
{\Ga(\frac{3}{2}-\frac{\ep}{2})}
\left[\de^\frac{\ep}{2}+\frac{\de^{1+\frac{\ep}{2}}}{2}
+\frac{3\de^{2+\frac{\ep}{2}}}{8}\right]
-\left[1+\frac{\de}{2-\ep}+\frac{(3-\ep)\de^2}{(2-\ep)(4-\ep)}
\right]\right\}
\nn\\
&&
+\od(\de^3,\de^{3+\frac{\ep}{2}})\,.
\eea
Our strategy for computing the divergent part of $\int_k\th_f(k^2)^n$
is now very simple:
Keep only powers of $\de$ in $\th_f(k^2)^n$ that are $\de^0$, $\de^1$
or $\de^2$ when $\ep\rightarrow 0$ (all higher powers of $\de$ lead
to convergent $k$ integrals).
Use
\beq
\label{intdelta}
\int_k\de^n
=
\int_k\left(\frac{4m^2}{k^2+4m^2}\right)^n
=
\frac{16m^4}{(4\pi)^2}
\left(\frac{m^2}{4\pi\mu^2}\right)^{-\frac{\ep}{2}}
\frac{2^{-\ep}\Ga(n-2+\frac{\ep}{2})}{\Ga(n)}
\eeq
to do the $k$ integration and expand in powers of $\ep$.
The terms with negative powers of $\ep$ give the divergent part.

Now let us check our new method for the two- and three-loop
integrals $I_1^{cc}$ and $I_2^{cc}$ and then use it to compute
the four-loop integral $I_3^{cc}$.
Expanding the two-loop integral
\beq
\label{i1rc}
I_1^{cc}
=\int_{pk}\frac{1}{[(k+p)^2+m^2](p^2+m^2)}
=I_{1A}^2
=\frac{m^4}{(4\pi)^4}\left(\frac{m^2}{4\pi\mu^2}\right)^{-\ep}
\Ga({\ts\frac{\ep}{2}-1})^2
\eeq
with $I_{1A}$ from (\ref{i1a}) in powers of $\ep$ gives the same
result as using our new method:
\bea
I_1^{cc}
&=&
\int_k\th(k^2)
=
\int_k[\th_d+\th_f(k^2)]
=
\int_k\th_f(k^2)
\nn\\
&=&
\frac{\Ga(\frac{\ep}{2})}{(4\pi)^2}
\left(\frac{m^2}{4\pi\mu^2}\right)^{-\frac{\ep}{2}}
\!\left\{\frac{\Ga(\frac{3}{2})\Ga(1-\frac{\ep}{2})}
{\Ga(\frac{3}{2}-\frac{\ep}{2})}
\int_k\left[\de^\frac{\ep}{2}{+}\frac{\de^{1+\frac{\ep}{2}}}{2}
{+}\frac{3\de^{2+\frac{\ep}{2}}}{8}\right]
-\int_k\left[1{+}\frac{\de}{2-\ep}
{+}\frac{(3-\ep)\de^2}{(2-\ep)(4-\ep)}
\right]\right\}
\nn\\
&&+\od(\ep^0)
\nn\\
&=&
\frac{16m^4\Ga(\frac{\ep}{2})}{(4\pi)^4}
\left(\frac{m^2}{4\pi\mu^2}\right)^{-\ep}2^{-\ep}
\left\{\frac{\Ga(\frac{3}{2})\Ga(1-\frac{\ep}{2})}
{\Ga(\frac{3}{2}-\frac{\ep}{2})}
\left[\frac{\Ga(\ep-2)}{\Ga(\frac{\ep}{2})}
+\frac{\Ga(\ep-1)}{2\Ga(1+\frac{\ep}{2})}
+\frac{3\Ga(\ep)}{8\Ga(2+\frac{\ep}{2})}\right]\right.
\nn\\
&&
\he\he\he\he\he\hh\left.
-\left[\frac{\Ga(\frac{\ep}{2}-1)}{2-\ep}
+\frac{(3-\ep)\Ga(\frac{\ep}{2})}{(2-\ep)(4-\ep)}\right]
\right\}+\od(\ep^0)
\nn\\
&=&
\frac{4m^4}{(4\pi)^4}
\left[\frac{1}{\ep^2}-\frac{1}{\ep}\left(\lnma-1\right)\right]
+\od(\ep^0)\,.
\eea
Of course, at two loops, this method seems awfully contrived.

Using (\ref{ircrecursion}), we get for the three-loop integral
\beq
\label{i2rc1}
I_2^{cc}=
\int_k\th_f(k^2)^2+2\th_d I_1^{cc}\,.
\eeq
Evaluating the divergent part of $\int_k\th_f(k^2)^2$ along the
lines of the strategy described above and using $\th_d=I_{1B}$ and
(\ref{i1b}) as well as $I_1^{cc}$ from (\ref{i1rc}), one recovers
(\ref{i2rcres}), as expected.

Having checked our method for two and three loops, we are now ready
to compute $I_3^{cc}$ and, in principle, circle-chain integrals
$I_n^{cc}$ to any number of loops $n+1$.
Using (\ref{ircrecursion}) to write
\beq
I_3^{cc}=\int_k\th_f(k^2)^3-3\th_d^2 I_1^{cc}+3\th_d I_2^{cc}\,,
\eeq
we can use our strategy to evaluate $\int_k\th_f(k^2)^3$.
Remembering that we do not keep a symbolic finite part of $I_1^{cc}$,
but evaluate it to the necessary order in $\ep$, the result is
\bea
\label{i3rc}
I_3^{cc}
&\!=\!&
\frac{m^4}{(4\pi)^8}
\left[
\frac{24}{\ep^4}
+\frac{1}{\ep^3}\left(-48\lnma+76\right)
+\frac{1}{\ep^2}\left(48\lnmb-152\lnma+134+12\ze(2)\right)
\right.
\nn\\
&&
\left.+\frac{1}{\ep}\left(
22\lnmc-55\lnmb+[47+30\ze(2)]\lnma-\frac{21}{2}-31\ze(2)+2\ze(3)
\right)\right]+\frac{6I_{2,f}^{cc}}{(4\pi)^2\ep}+I_{3,f}^{cc}\,,
\nn\\
\eea
where $I_{3,f}^{cc}=\od(\ep^0)$.

\section{Hypergeometric Function}
%================================
\label{hg}

Here are some formulas for Gauss's hypergeometric function
$F(a,b;c;z)$, used for the computation of the circle-chain
integrals $I_n^{cc}$.
The formulas are taken directly or slightly modified from
\cite{GrRh}.

$F(\al,\be;\ga;z)$ is defined by the series
\beq
\label{hgex}
F(\al,\be;\ga;z)=
1+\frac{\al\be}{\ga\cdot 1}z
+\frac{\al(\al+1)\be(\be+1)}{\ga(\ga+1)\cdot 1\cdot 2}z^2
+\frac{\al(\al+1)(\al+2)\be(\be+1)(\be+2)}{\ga(\ga+1)(\ga+2)
\cdot 1\cdot 2\cdot 3}z^3
+\cdots\,.
\eeq
A relevant integral is
\beq
\label{hgint}
\int_0^1\frac{dx\,x^\mu}{(1+a x)^\nu}
=
\frac{1}{\mu+1}F(\nu,\mu+1;\mu+2;-a)\,.
\eeq
A transformation formula is
\bea
\label{hgtransform}
F(\al,\be;\ga;1-z)
&=&
\Ga(\ga)\left[
\frac{\Ga(\ga-\al-\be)}{\Ga(\ga-\al)\Ga(\ga-\be)}
F(\al,\be;\al+\be-\ga+1;z)\right.
\nn\\
&&
\he\hh +z^{\ga-\al-\be}
\left.\frac{\Ga(\al+\be-\ga)}{\Ga(\al)\Ga(\be)}
F(\ga-\al,\ga-\be;\ga-\al-\be+1;z)
\right]\,.
\eea
A representation of an elementary function:
\beq
\label{hgelrep}
F(\al,\mu;\al;z)=F(\mu,\be;\be;z)=(1-z)^{-\mu}\,.
\eeq

\setlength{\baselineskip}{14pt}

\end{document}